\documentclass{amsart}
\usepackage{amscd}
\usepackage{thmdefs}
\usepackage{graphicx}

\input{tcilatex}
\input tcilatex

\begin{document}
\title{Discrete and oscillatory Matrix Models in Chern-Simons theory}
\author{Sebastian de Haro}
\address{Max-Planck f\"{u}r Gravitationsphysik, Albert-Einstein-Institut, 14476 Golm,
Germany\\
sdh@aei.mpg.de}
\author{Miguel Tierz}
\address{Institut d'Estudis Espacials de Catalunya (IEEC), Campus UAB, Fac. Ciencies,
Torre C5-Par-2a pl, E-08193 Bellaterra (Barcelona), Spain\\
tierz@ieec.fcr.es}
\address{The Open University, Applied Mathematics Department., Milton Keynes, MK7
6AA, UK\\
M.Tierz@open.ac.uk}
\date{}

\begin{abstract}
We derive discrete and oscillatory Chern-Simons matrix models. The method is
based on fundamental properties of the associated orthogonal polynomials. As
an application, we show that the discrete model allows to prove and extend
the recently found relationship between Chern-Simons theory and $q$-deformed
2dYM. In addition, the equivalence of the Chern-Simons matrix models gives a
complementary view on the equivalence of effective superpotentials in $%
\mathcal{N}=1$ gauge theories.
\end{abstract}
\maketitle

\section{\protect\smallskip Introduction}

Since the seventies, when random matrix theory \cite{Mehta} was successfully
employed in gauge theory \cite{Brezin:1977sv}, matrix models have played a
rather remarkable role in gauge theory (see \cite
{DiFrancesco:1993nw,Marino:2004eq} for reviews). In this paper, we shall
focus on the matrix models that appear in Chern-Simons theory \cite
{Marino:2002fk,Tierz} and also discuss the recently found connection with
2dYM \cite{deHaro:2004uz,Aganagic:2004js} as well as $\mathcal{N}=1$
supersymmetric gauge theories. More precisely, with recent work that
discusses the equivalence of effective superpotentials in these theories 
\cite{eq1,eq2,eq3}.

Let us briefly recall very basic facts of Chern-Simons gauge theory: In \cite
{cs}, Witten considered a topological gauge theory for a connection on an
arbitrary three-manifold $M,$ based on the Chern-Simons action: 
\begin{equation}
S_{\mathrm{CS}}(A)={\frac{k}{4\pi }}\int_{M}\mathrm{Tr}(A\wedge dA+{\frac{2}{%
3}}A\wedge A\wedge A),  \label{cs}
\end{equation}
with $k$ an integer number. One of the most important aspects of
Chern-Simons theory is that it provides a physical approach to three
dimensional topology. In particular, it gives three-manifold invariants and
knot invariants. For example, the partition function,

\begin{equation}
Z_{k}(M)=\int \mathcal{D}A\mathrm{e}^{iS_{\mathrm{CS}}(A)},  \label{wrt}
\end{equation}
delivers a topological invariant of $M$, the so-called
Reshetikhin-Turaev-Witten invariant. Recent reviews are \cite
{Marino:2004eq,Marino:2004uf}.

As reviewed in detail in \cite{Marino:2004uf}, a great deal of interest has
been recently focused on the fact that Chern-Simons theory provides large $N 
$ duals of topological strings. This connection between Chern-Simons theory
and topological strings was already pointed out by Witten \cite{wittenopen}
(see also \cite{Periwal}), and then considerably extended in \cite
{Gopakumar:1998ki}.

Interestingly enough, Mari\~{n}o has found a tight connection between
Chern-Simons theory and random matrix models \cite{Marino:2002fk}. More
precisely, building upon \cite{Rozseifert}, in \cite{Marino:2002fk} it is
shown that the partition function of Chern-Simons theory on $S^{3}$ with
gauge group $U(N)$ is given by the partition function of the following
random matrix model: 
\begin{equation}
Z=\frac{\mathrm{e}^{-\frac{g_{s}}{12}N\left( N^{2}-1\right) }}{N!}\int
\prod_{i=1}^{N}\mathrm{e}^{-u_{i}^{2}/2g_{s}}\prod_{i<j}\left( 2\sinh \frac{%
u_{i}-u_{j}}{2}\right) ^{2}\frac{du_{i}}{2\pi },  \label{sinh}
\end{equation}
From the point of view of topological strings, this describes open
topological $A$ strings on $T^{*}\emph{S}^{3}$ with $N$ branes wrapping $%
S^{3}$ \cite{Marino:2002fk}. Chern-Simons matrix models have been further
considered in \cite{Tierz} and \cite{Aga}-\cite{deHaro:2004wn}. Most of
these works focus on the relevance to topological strings. In \cite{Tierz},
the emphasis is on exact solutions and on the special features of the matrix
models. In this paper we shall be further developing these results, while
making some connections, at the end, with matrix models in $\mathcal{N}=1$
suspersymmetric gauge theories.

It should be stressed that the result $\left( \ref{sinh}\right) $ in \cite
{Marino:2002fk} is just a particular case of more general models and our
results will apply to them as well. Explicit expressions for them will be
given in the next section.

As shown in \cite{Tierz}, the Stieltjes-Wigert polynomials, a member of the $%
q$-deformed orthogonal polynomials family \cite{Koe}, allows to compute, in
exact fashion, quantities associated to the matrix model. In the
computation, the $q$-parameter of the polynomials turns out to be naturally
identified with the $q$-parameter of the quantum group invariants associated
to the Chern-Simons theory. In \cite{Tierz}, the focus was on $U(N)$
Chern-Simons theory on $S^{3}$, whose matrix model is also the
Stieltjes-Wigert ensemble: 
\begin{equation}
Z=\int [dM]\mathrm{e}^{-{\frac{1}{2g_{s}}}{\mathrm{Tr}}(\log M)^{2}}~.
\end{equation}

A special property of any of these type of matrix models is that there are
infinitely many models that lead to the same partition function \cite{Tierz}%
. This is a courtesy of their very weakly confining potentials. We shall
extend this result, essentially by exploiting results from the moment
problem \cite{Sti,Sim}, as in \cite{Tierz}, in a more explicit way. In some
cases, it just suffices to consider well-known facts of orthogonal
polynomials with undetermined weight \cite{Sim}. In others, the ones
corresponding to the oscillatory/periodic models, one should further develop
-in a simple and natural way-, the results of the moment problem to fully
understand the extent of the equivalence between the models.

To summarize, as a first result we have the existence of Chern-Simons matrix
models with a discrete weight, which essentially turns out to be a
discretization of the weight of the usual models \cite{Marino:2002fk,Tierz}.
That is to say, the continuous models in \cite{Marino:2002fk,Tierz} can be
castled into an homogenous and exponential lattice, respectively. As we
shall see, this discrete model is precisely the $q$-deformed 2dYM partition
function. The advantage of our method is that it can be applied to any
Chern-Simons matrix model. Then, we shall deal with the continuous case, and
show the manifold possibilities, including losing differentiability of the
weight function. From a mathematical point of view, it is remarkable that
the potential may be so wild even to reach fractal behavior and still leave
the partition function (which in addition is a Chern-Simons partition
function) unchanged. Interpretations in terms of gauge theories and quantum
groups are discussed in some instances, and we show that the equivalence of
potentials of the (generalized) Chern-Simons matrix models introduced, leads
to a complementary view of the equivalence of effective superpotentials in $%
\mathcal{N}=1$ supersymmetric gauge theories \cite{eq1,eq2,eq3}. In the last
section, we comment on the possible physical relevance and some further
avenues of research are suggested.

\section{Generalized Chern-Simons matrix models}

The idea, already introduced in \cite{Tierz}, is that Chern-Simons
quantities can be obtained from a very precise coarse-grained knowledge of
infinitely many different distributions, not just the log-normal. In matrix
model language: there are infinitely many, equivalent, matrix models.
Remarkable particular cases are:

1) Discrete matrix model. The discretization of the models in \cite
{Marino:2002fk,Tierz} is deeply related to the quantum group symmetry of
Chern-Simons theory, and leads directly to 2dYM \cite
{deHaro:2004uz,Aganagic:2004js}.

2) The continuous and periodic models. The main feature is the presence of
log-periodic \cite{long,Sor}\footnote{%
This behavior is typical of models with a discrete-scale invariance \cite
{Sor}. It appears in rigorous formulations of fractal geometry \cite{long}
and in several condensed-matter physics applications \cite{Sor}. Connections
with quantum group symmetries are discussed in \cite{Tierz2}. Recent
physical applications involving this behavior include limit-cycles in RG
flows (see \cite{limit} for example) and critical phenomena in gravitational
collapse \cite{critical}.} and periodic behavior.

These examples underline to what extent the matrix models can be deformed
while preserving the Chern-Simons information.

\subsection{Preliminaries}

For future use, we quote here the moments of the weight function associated
with the Stieltjes-Wigert polynomials: 
\begin{equation}
{\frac{k}{\sqrt{\pi }}}\int_{0}^{\infty }{\mbox{d}}x\,x^{n}\mathrm{e}%
^{-k^{2}(\log x)^{2}}=\mathrm{e}^{(n+1)^{2}/4k^{2}}.  \label{moments0}
\end{equation}
In what follows we will set $q=\mathrm{e}^{-1/2k^{2}}$. Notice that the
above formula is valid as long as ${\mbox{Re}}\,k^{2}\geq 0$. For the
applications in Chern-Simons theory, we will be interested in the two cases $%
{\mbox{Re}}\,k^{2}=0$ and ${\mbox{Im}}\,k^{2}=0$.

We now consider the set of weights that satisfy the functional equation: 
\begin{equation}
w(qx)=\sqrt{q}\,x\,w(x),  \label{func}
\end{equation}
it is easy to see that $w$ also satisfies 
\begin{equation}
w(xq^{n})=q^{n^{2}/2}x\,w(x).
\end{equation}
Let us normalize $w$ such that 
\begin{equation}
q^{-1/4}\int_{0}^{\infty }{\mbox{d}}x\,w(x)=1~.
\end{equation}
Applying the proof of proposition 2.1 of \cite{Christ} to this weight, it
now follows that it has the same moments as the log-normal weight: 
\begin{equation}
\int_{0}^{\infty }{\mbox{d}}x\,x^{n}w(x)=q^{-(n+1)^{2}/2}~.  \label{moments}
\end{equation}
The log-normal distribution (\ref{moments0}) is a distinguished case of an
undetermined moment problem and a basic result from the theory of moments is
that the set of solutions of an indeterminate moment problem contains
discrete measures as well as absolutely continuous measures. Indeed, it was
Stieltjes in his seminal work \cite{Sti}, the first to show that all the
functions in the following family: 
\begin{equation}
f_{\vartheta }\left( x\right) =\mathrm{e}^{-\log ^{2}x}\left( 1+\vartheta
\sin \left( 2\pi \log x\right) \right) ,\quad \text{with}\ \vartheta \in
\left[ -1,1\right] ,
\end{equation}
have the same moments, the log-normal moments. Thus, the parameter $%
\vartheta $ does not play any role regarding integer moments and the
infinitely many functions in the family $g_{\vartheta }\left( x\right) $ all
have the same integer moments (and consequently, the same orthogonal
polynomials). This case corresponds to $k=1$. More generally, ${}$for an
arbitrary $k$, the function is then: 
\begin{equation}
f_{\vartheta }\left( x\right) =\mathrm{e}^{-k^{2}\log ^{2}x}\left(
1+\vartheta \sin \left( 2\pi \frac{\log x}{\log q}\right) \right) ,\quad 
\text{with}\ \vartheta \in \left[ -1,1\right] ,  \label{Stifam}
\end{equation}
and with $q\equiv \mathrm{e}^{-1/2k^{2}}$ as above.

We consider one more class of weight functions, namely the discrete ones 
\cite{Christ,Chi}: 
\begin{equation}
w(x)={\frac{1}{\sqrt{q}M(c)}}\sum_{n=-\infty }^{\infty }c^{n}q^{{\frac{n^{2}%
}{2}}+n}\delta (x-cq^{n}),  \label{discreteweight}
\end{equation}
where 
\begin{equation}
M(c)=(-cq\sqrt{q},-\sqrt{q}/c,q;q)_{\infty }~.
\end{equation}
From now on we will take $q$ to be real and $c$ an arbitrary real positive
constant, $c>0$. It is not hard to see that $w$ again has the same moments (%
\ref{moments}): one just uses the Jacobi triple product identity 
\begin{equation}
\sum_{n=-\infty }^{\infty }(-1)^{n}q^{\binom{n}{2}}x^{n}=(x,q/x,q;q)_{\infty
},
\end{equation}
and translates the sum.

\subsection{Generic description}

In fact, it is possible to have a more general characterization of the
solutions of the log-normal moment problem. Following \cite{Chi,Christ},
consider two generic weight functions, $\omega _{1}\left( x\right) $ and $%
\omega _{2}\left( x\right) $ satisfying the functional equation $\left( \ref
{func}\right) $. Since the functional equation determines the log-normal
moments, these generic functions possess the log-normal moments. Consider
the quotient: $g\left( x\right) =$ $\omega _{1}\left( x\right) /$ $\omega
_{2}\left( x\right) ,$ which leads to a function that satisfies a functional
equation: 
\begin{equation}
g\left( x\right) =g\left( qx\right) ,  \label{func3}
\end{equation}
and thus, two generic functions with log-normal moments differ by a function
that satisfies $\left( \ref{func3}\right) $. This type of function can be
naturally named, as in \cite{Chi,Christ}, $q$-periodic, and it is manifest
that any logarithmic oscillatory term (with the proper frequency, as in $%
\left( \ref{Stifam}\right) $ for example) satisfies this property. This type
of function is very natural when $q$ deformations are present.

Thus, any integral constructed from the moments with any of the above
weights will give the same result. We can now apply this to matrix models
where we have $N$ such integrals. Consider the following matrix model: 
\begin{equation}
Z=\int_{0}^{\infty }{\mbox{d}}x_{1}\ldots \int_{0}^{\infty }{\mbox{d}}x_{N}\,%
\mathrm{e}^{-{\frac{1}{2g_{s}}}\sum_{i=1}^{N}\log
^{2}x_{i}}\prod_{k<l}(x_{k}-x_{l})^{2}\,P(x_{1},\ldots ,x_{N}),  \label{mm}
\end{equation}
where $P$ has a finite Laurent expansion in the $x$'s. We write it as
follows: 
\begin{equation}
P(x_{1},\ldots ,x_{N})=\sum_{n_{1}\ldots n_{N}}p_{n_{1}\ldots
n_{N}}x_{1}^{n_{1}}\ldots x_{N}^{n_{N}},
\end{equation}
for some coefficients $p_{n_{1}\ldots n_{N}}$, where the $n_{i}$'s can be
positive or negative integers. Notice that $P$ is a symmetric Laurent series
in the $x_{i}$'s, since an antisymmetric piece in any two $x_{i}$'s would
not contribute to the integral. We can expand $P$ in a basis of symmetric
polynomials or, alternatively, a basis of Schur functions.

It is now obvious that (\ref{mm}) inherits the invariance under deformation
of the weights as discussed above. The Vandermonde interaction itself is a
polynomial of degree $N(N-1)$ in $x$, so the integrand factorizes into a
product of integrals over the $x_{i}$'s, and the weights can be deformed for
each integral. Thus, we get that 
\begin{eqnarray}
Z &=&\int_{0}^{\infty }{\mbox{d}}x_{1}\ldots \int_{0}^{\infty }{\mbox{d}}%
x_{N}\,\mathrm{e}^{-{\frac{1}{2g_{s}}}\sum_{i=1}^{N}\log
^{2}x_{i}}\prod_{k<l}(x_{k}-x_{l})^{2}\,P(x_{1},\ldots ,x_{N})  \notag \\
&=&\int_{0}^{\infty }{\mbox{d}}x_{1}\ldots \int_{0}^{\infty }{\mbox{d}}%
x_{N}\,w(x_{1})\ldots w(x_{N})\,\prod_{k<l}(x_{k}-x_{l})^{2}\,P(x_{1},\ldots
,x_{N}),
\end{eqnarray}
where $w(x)$ is any of the above weights, whether continuous or discrete.

Therefore, the following matrix model is also invariant under the same
deformations: 
\begin{equation}
Z=\int [dM]\mathrm{e}^{-{\frac{1}{2g_{s}}}{\mathrm{Tr}}(\log M)^{2}}\,f(M),
\end{equation}
where $f(M)$ has a finite Laurent expansion in the moments ${\mbox{Tr}}%
\,M^{n}$, and $M$ is a Hermitian matrix.

We now apply the above to Chern-Simons theory. Consider the partition
function of Chern-Simons theory on a Seifert space $M=X({\frac{p_{1}}{q_{1}}}%
,\ldots ,{\frac{p_{n}}{q_{n}}})$. This is obtained by doing surgery on a
link in $S^{3}$ with $n+1$ components, out of which $n$ are parallel,
unlinked unknots, and one has link number $1$ with each of the $n$ unknots.
The surgery data are $p_{j}/q_{j}$ for the unlinked unknots, $j=1,\ldots ,n$%
, and 0 for the last component. The partition function is\footnote{%
See the Appendix for details on the notation} \cite{Marino:2002fk}: 
\begin{eqnarray}
\mathbf{Z}_{\mathrm{CS}}(M) &=&{\frac{(-1)^{|\Delta _{+}|}}{|\mathcal{W}%
|\,(2\pi i)^{r}}}\Biggl( {\frac{\mathrm{Vol}\,\Lambda _{\mathrm{w}}}{\mathrm{%
Vol}\,\Lambda _{\mathrm{r}}}}\Biggr){\frac{[\mathrm{sign}(P)]^{|\Delta _{+}|}%
}{|P|^{r/2}}}\mathrm{e}^{{\frac{\pi id}{4}}\mathrm{sign}(H/P)-{\frac{\pi idy%
}{12l}}\phi }  \label{20} \\
&\times &\sum_{t\in \Lambda _{\mathrm{r}}/H\Lambda _{\mathrm{r}}}\int d\beta
\,\mathrm{e}^{-{\beta ^{2}/2g_{s}}-lt\cdot \beta }{\frac{\prod_{i=1}^{n}%
\prod_{\alpha >0}2\sinh {\frac{\beta \cdot \alpha }{2p_{i}}}}{\prod_{\alpha
>0}\left( 2\sinh {\frac{\beta \cdot \alpha }{2}}\right) ^{n-2}}}~.  \notag
\end{eqnarray}
This expression gives the contribution of the reducible flat connections to
the partition functions. Recall that for both $S^{3}$ and lens spaces this
amounts to the exact partition function. The case $n=0$ corresponds to the
three-sphere $S^{3}$ that leads to $\left( \ref{sinh}\right) .$ Thus, for
the case of $U(N)$, and focusing on a particular sector of flat connections,
we get the following matrix model: 
\begin{equation}
Z_{\mathrm{CS}}(M)=\prod_{i=1}^{N}\int_{-\infty }^{\infty }{\mbox{d}}y_{i}\,%
\mathrm{e}^{-y_{i}^{2}/2g_{s}-lt_{i}y_{i}}{\frac{\prod_{j=1}^{n}\prod_{k<l}2%
\sinh {\frac{y_{k}-y_{l}}{2p_{j}}}}{\prod_{k<l}\left( 2\sinh {\frac{%
y_{k}-y_{l}}{2}}\right) ^{n-2}}}~.
\end{equation}
We used a different letter to remind ourselves that to obtain the full
partition function one needs to include the constant prefactor and sum over $%
t$. Notice that because of the integrals, the integrand in the above
expression is automatically symmetric in the $t_{i}$'s.

All we have to do in order to show invariance under deformations of the
weight is to show that this is of the form (\ref{mm}). But this should be
clear. As in \cite{Tierz}, we perform a coordinate transformation 
\begin{equation}
y_{i}=a\log x_{i}+b,  \label{cotrafo}
\end{equation}
for some constants $a$ and $b$. We get: 
\begin{eqnarray}
Z_{\mathrm{CS}}(M) &=&a^{N}\mathrm{e}^{-{\frac{Nb^{2}}{2g_{s}}}%
-bl\sum_{i=1}^{N}t_{i}}\int_{0}^{\infty }{\mbox{d}}x_{1}\ldots
\int_{0}^{\infty }{\mbox{d}}x_{N}\,\mathrm{e}^{-{\frac{a^{2}}{2g_{s}}}%
\sum_{i=1}^{N}\log ^{2}x_{i}}(x_{1}\ldots x_{N})^{-d}  \notag
\label{seifert} \\
&&\times \prod_{i=1}^{N}{\frac{1}{x_{i}^{alt_{i}}}}\prod_{k<l}\left(
(x_{k}^{a}-x_{l}^{a})^{2-n}\prod_{i=1}^{n}(x_{k}^{a/p_{i}}-x_{l}^{a/p_{i}})%
\right) ,
\end{eqnarray}
where 
\begin{equation}
d=1+{\frac{ab}{g_{s}}}+{\frac{1}{2}}\,a(N-1)(\sum_{i=1}^{n}{\frac{1}{p_{i}}}%
+2-n)~.
\end{equation}
Obviously, a sufficient condition for this to be of the form (\ref{mm}) is
to choose $a=\tilde{a}\,p_{1}\ldots p_{n}$, $b=\tilde{c}\,g_{s}$, where $%
\tilde{a},\tilde{b}\in {\mathbb{Z}}$. Thus, all these models can be deformed
in the way shown above. Basically, any matrix model with an integrand where
the periodicity is broken by a Gaussian weight, will have this deformation
property. This includes all of the known Chern-Simons matrix models, the
most general one being that for torus link invariants \cite{Marcos}. We will
discuss this elsewhere \cite{inprep}.

\subsection{Discrete model and 2d qYM}

In the early 70's, Chihara employed the functional equation $\left( \ref
{func}\right) $ to show the existence of an infinite number of discrete
measures with the same log-normal moments \cite{Chi}. An example is $\left( 
\ref{discreteweight}\right) $, discussed above. This case is remarkable,
since it is completely explicit. Furthermore, it not only leads to a
discrete Chern-Simons but it also turns out that this model is immediately
relevant to the correspondence between Chern-Simons theory and 2D Yang-Mills 
\cite{deHaro:2004uz} as we shall see at the end of this section.

Note that the moments of $\left( \ref{discreteweight}\right) $ do not depend
on the parameter $c$. Thus, as in the case of the continuous functions, this
example already contains infinitely many functions with the same moments.
The discrete case leads, for $U(N)$ and $S^{3}$, to the following
Chern-Simons matrix model :

\begin{eqnarray}
Z &=&C_{N}\int \prod_{i=1}^{N}dx_{i}\sum_{n=-\infty }^{\infty
}c^{n}q^{n+n^{2}/2}\delta \left( x-cq^{n}\right) \prod_{i<j}\left(
x_{i}-x_{j}\right) ^{2}  \label{disc} \\
= &&\widetilde{C}_{N}\sum_{n_{1}=-\infty }^{\infty }\cdot \cdot \cdot
\sum_{n_{N}=-\infty }^{\infty
}\prod_{i=1}^{N}c^{n_{i}}q^{n_{i}^{2}+n_{i}/2}\prod_{i<k}\left(
q^{n_{j}}-q^{n_{k}}\right) ^{2}.  \notag
\end{eqnarray}

At this point, it is worth to recall the works on combinatorial quantization 
\cite{Schom,Schom2}, whose aim is a mathematically careful quantization of
pure Chern-Simons theory. These works \textit{simulate} Chern-Simons theory
on a lattice, in such a way that partition functions and correlators of the
lattice model coincide with those of the continuous model. Interestingly
enough, in \cite{Schom} it is shown that one can also choose the point of
view of constructing noncommutative gauge fields starting from some symmetry
algebra placed at the lattice sites. It might be a quantum symmetry algebra,
but one can also choose another symmetry algebra. In \cite{Schom}, they
choose the gauge symmetry algebra. Doing this, it is found that the gauge
theory is indeed reconstructed if the symmetry algebra is endowed with a
co-multiplication. That is to say, that it is extended to a Hopf algebra,
essentially. This is a remarkable check of the role of the quantum group
symmetry in the discretization.

The previous expression was a discretization of the Stieltjes-Wigert
ensemble, the discretization of $\left( \ref{sinh}\right) $ would be:

\begin{equation}
Z=\sum_{u_{1},...,u_{N}=-\infty }^{\infty }\mathrm{e}^{-g_{s}%
\sum_{i=1}^{N}u_{i}^{2}}\prod_{i<j}\sinh ^{2}\left( u_{i}-u_{j}\right)~.
\label{sinhdisc}
\end{equation}

Note also that, as one can easily see from the observation of the value of
the coefficients at each point of the lattice, we do not only have a
discretization of the log-normal or the Gaussian, but there is also a sort
of modular transformation -like in Poisson resummation for example- since
the weights are of the type $q^{n^{2}}=\mathrm{e}^{-g_{s}n^{2}},$ in
contrast to the $\mathrm{e}^{-n^{2}/g_{s}}$ that would come from a direct
discretization of the original matrix models$.$ This also seems to be
analogous to what happens in combinatorial quantization, where the symmetry
at each site of the lattice is a quantum group symmetry.

The formula (\ref{sinhdisc}), coming from the partition function of
Chern-Simons on $S^{3}$, is easily seen to give the partition function of
2dYM on the cylinder with trivial holonomies around the two endpoints of the
cylinder \cite{Gross:1994ub}. It is also the partition function of $q$%
-deformed 2dYM on $S^{2}$ \cite{Aganagic:2004js}. As explained in \cite
{deHaro:2004uz,deHaro:2004wn}, both facts agree since these partition
functions are the same. Thus, our discretization directly proves the
equivalence between the Chern-Simons matrix model and 2dYM, including the
numerical prefactors, which we did not keep track of in (\ref{sinhdisc}).
This applies to the more general cases worked out in \cite{deHaro:2004uz}
and \cite{Aganagic:2004js}, and in fact it can be generalized to all the
cases where discretization applies (for example, the Seifert homology
spheres discussed above). It can also be used in the opposite direction,
that is, one starts with a discrete model and rewrites it as a continuous
matrix model. This can be helpful to compute large $N$ limits of such
models, since the large $N$ techniques are much better understood in the
continuous case. We will discuss this in more detail elsewhere \cite{inprep}.


\section{Oscillatory behavior in Chern-Simons matrix models}

\subsection{Connection with complex dimensions and fractal behavior}

The connection with complex dimensions \cite{long,Sor} can now be easily
obtained. We consider the Mellin transform of the $q$-periodic function $%
\left( \ref{func3}\right) $: 
\begin{equation}
h(s)\equiv \int_{0}^{\infty }\widetilde{\omega }\left( x\right) x^{s}dx.
\end{equation}
Taking into account the following property of Mellin transforms: 
\begin{equation}
\int_{0}^{\infty }\widetilde{\omega }\left( qx\right) x^{s}dx=q^{-s}h(s),
\end{equation}
and considering the $q$-periodic property $\left( \ref{func3}\right) :$%
\begin{equation}
q^{-s}h(s)=h\left( s\right) \Rightarrow s_{n}=\frac{2\pi in}{\log q},\quad
n\in {\mathbb{Z}}~.
\end{equation}
Thus, the Mellin transform of a $q$ periodic function contains infinitely
many complex poles. From standard results in Mellin asymptotics, this
implies that the $q$ periodic function may be fractal. This depends on the
precise form of the function. Namely, on the residue of the poles. This is a
usual situation in Mellin asymptotics with complex poles. If the residues
decay fast enough with $n,$ then the function will be differentiable.

Another way to arrive at the same conclusion as above, is to notice that $%
1+\vartheta \sin \left( 2\pi \frac{\log x}{\log q}\right) $ in $\left( \ref
{Stifam}\right) $ is just one possible deformation among many. The parameter 
$\vartheta $ takes its possible values in $\left[ -1,1\right] $ just to
ensure that the resulting function is positive definite everywhere. Of
course, the oscillatory term can be easily generalized. For example, one can
consider: 
\begin{equation}
f\left( x\right) =\mathrm{e}^{-k^{2}\log ^{2}x}\left(
1+\sum_{n=1}^{m}a_{n}\sin \left( 2\pi n\frac{\log x}{\log q}\right) \right) ,
\label{logper}
\end{equation}
and the integer moments keep their log-normal value.

\subsection{Periodic/Self-similar matrix models}

Note that in $\left( \ref{logper}\right) $ we are still limiting the finest
scale possible, since the sum stops at $n=m$ and thus the function is not a
genuine fractal, albeit the oscillatory pattern is certainly much richer
than in $\left( \ref{Stifam}\right) $. But we can consider a full Fourier
series and take $m\rightarrow \infty $ as long as we restrict the
coefficients in order to ensure positiveness of the function. This
restriction is actually very mild. Consider for example, $a_{n}=n^{-\gamma }$
with $\gamma >1;$ then: 
\begin{equation}
f\left( x\right) =\mathrm{e}^{-k^{2}\log ^{2}x}\left( 1+\lambda
\sum_{n=1}^{\infty }\frac{1}{n^{\gamma }}\sin \left( 2\pi b_{n}\frac{\log x}{%
\log q}\right) \right) ,  \label{general}
\end{equation}
is positive definite for a low enough value of the parameter $\lambda $. One
of these functions looks as:

\begin{figure}[htp]
\centering
\includegraphics{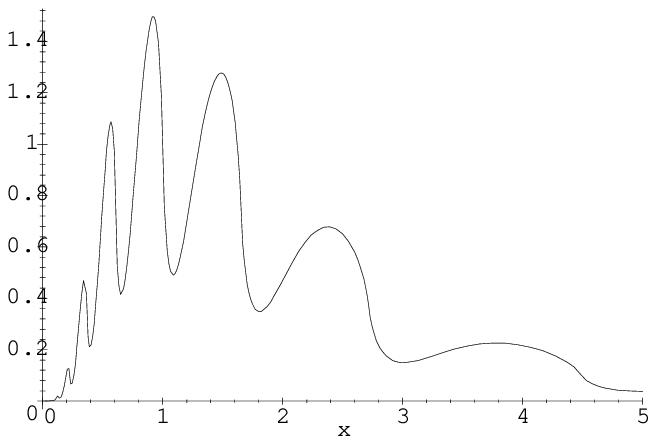}
\caption{$f(x)$ in (\ref{general}) with $a_{n}=n^{-2},b_{n}=n,k=1,\lambda
=0.5,m=200$}
\end{figure}

It is plain that for some low enough values of $\gamma ,$ the resulting
function is fractal \cite{Mand}. Namely, continuous but differentiable
nowhere. An example of such a function is plotted in Figure 2.

\begin{figure}[htp]
\centering
\includegraphics{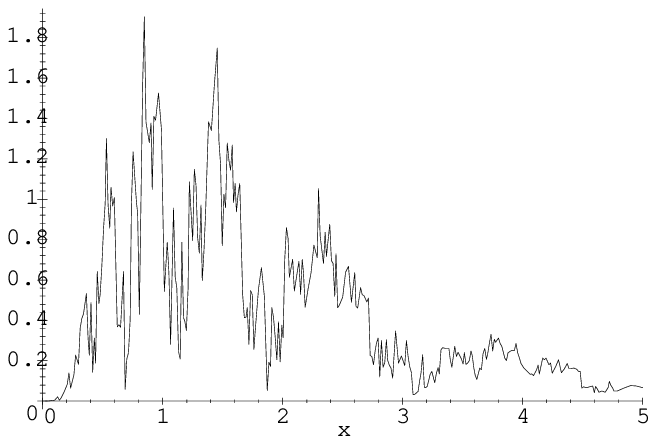}
\caption{$f(x)$ in (\ref{general}) with $a_{n}=n^{-1},b_{n}=n^{2},k=1,%
\lambda=0.5, m=50$}
\end{figure}

Therefore, we have a family of non-polynomial potentials, all of them
equivalent in the sense that they give rise to the same partition functions.
The other possible relevant quantities, such as the density of states and
correlation functions are trivially related \cite{Tierz}.

Going back to the original Chern-Simons matrix model, we are lead to
ensembles such as, for example: 
\begin{equation}
Z=\int \prod_{i=1}^{N}\mathrm{e}^{-\frac{u_{i}^{2}}{2g_{s}}}\left( 1+\lambda
\sum_{n=1}^{\infty }a_{n}\sin \left( 2\pi b_{n}\frac{u_{i}+Ng_{s}}{\log q}%
\right) \right) du_{i}\prod_{i<j}\left( 2\sinh \frac{u_{i}-u_{j}}{2}\right)
^{2},
\end{equation}
with suitable $a_{n}$ and $b_{n}$. At the level of the confining potential
we have: 
\begin{equation}
V\left( u\right) =u^{2}-\ln \left( 1+\lambda \sum_{n=1}^{\infty }a_{n}\sin
\left( 2\pi b_{n}\frac{u+Ng_{s}}{\log q}\right) \right) ,  \label{potential}
\end{equation}

Since $V\left( x\right) =-\ln \omega \left( x\right) ,$ the derivative of
the potential is the logarithmic derivative of the weight function $%
V^{\prime }\left( x\right) =-\frac{\omega ^{\prime }\left( x\right) }{\omega
\left( x\right) }$ and thus, non-differentiability of $\omega ^{\prime
}\left( x\right) $ implies non-differentiability of $V^{\prime }\left(
x\right) ,$ since $\omega \left( x\right) $ is a continuous function.
Therefore, the potential itself may be a fractal. This possibility depends
on the choice of the coefficients $a_{n}$ and $b_{n}$ as the previous
figures clearly show. A more detailed analysis follows from results in
Fourier series \cite{Mand}. While we shall discuss this mathematical
property elsewhere, recall that if we have a Fourier series such as $f\left(
u\right) =\sum_{m}a_{m}\exp \left( imu\right) ,$ where $a_{m}$ have random
or pseudorandom phases, then, if the power spectrum $\left| a_{m}\right|
^{2} $ has the asymptotic form: 
\begin{equation}
\left| a_{m}\right| ^{2}\sim \left| m\right| ^{-\beta }\quad \text{as }%
\left| m\right| \rightarrow \infty ,\text{ where }1<\beta \leq 3,
\end{equation}
the graphs of Re$f$ and Im$f$ are continuous but non-differentiable, with
fractal dimension: 
\begin{equation}
D_{f}=\frac{1}{2}\left( 5-\beta \right) .
\end{equation}

\subsection{Oscillatory behavior of the bare model: density of states.}

The way oscillatory behavior appears in the matrix models may seem
surprising. Nevertheless, discrete scale invariance is known to be related
to $q$-deformations with $q$ real \cite{Tierz2}, so its appearance in
Chern-Simons matrix models is natural in fact. Furthermore, the oscillatory
terms seem to be intimately related to an $SL(2,{\mathbb{Z}})$ invariance.
We shall discuss this elsewhere, but at least we can readily show now that
oscillatory behavior is also a feature even in the original \textit{bare }%
model (\ref{sinh}). For this, we have just to consider the famous expression
for the density of states of a Hermitian matrix model \cite{Mehta}: 
\begin{equation}
\rho \left( x\right) =\omega \left( x\right) \sum_{n=0}^{N-1}P_{n}^{2}\left(
x\right) ,  \label{dens}
\end{equation}
valid for any $N$. Some features of this quantity have already been studied
as they are useful in the context of topological strings \cite{Marino:2004eq}%
. However, in this case, one is interested in the perturbative expansion in $%
N$. This leads one to consider the 't Hooft limit where $\lambda =Ng_{s}$ is
kept constant while $N\rightarrow \infty $. Since this corresponds to $q\sim
1$, the limit where the $q$-matrix model approaches a classical one, we lose
the oscillations in (\ref{dens}), as we shall see. On the other hand, to
give a generic expression (e.g. the analogous of the semi-circle law for the
Hermitian Gaussian model) is also not very useful. The problem is that there
are apparently no analogues of the Plancherel-Rotach asymptotic formulas 
\cite{Mehta} and therefore any expression is rather cumbersome. But just to
get a feeling of the behavior of the density of states for arbitrary $q$, we
can carry -employing \textit{Mathematica}, for example- exact analytic
computations for various values of $N$ and $q$, to get an idea of the
beautiful behavior of the density of states. More precisely, we employ (\ref
{dens}) with: 
\begin{eqnarray}
\text{ }P_{n}\left( x\right) &=&\left( -1\right)
^{n}q^{n/2+1/4}\prod_{j=1}^{n}\left( 1-q^{j}\right) ^{-1/2}\sum_{\upsilon
=0}^{n}\binom{n}{\upsilon }_{q}q^{\upsilon ^{2}}\left( -q^{1/2}x\right)
^{\upsilon },  \label{SW} \\
\omega \left( x\right) &=&\mathrm{e}^{-k^{2}\log ^{2}x}\text{ with }\binom{n%
}{\upsilon }_{q}=\frac{\prod_{i=0}^{\upsilon -1}\left( 1-q^{n-i}\right) }{%
\prod_{i=0}^{\upsilon -1}\left( 1-q^{i+1}\right) }.  \notag
\end{eqnarray}
That is, the log-normal weight function and the Stieltjes-Wigert polynomials 
\cite{Christ}. We also go back to the original variables, employing $x=%
\mathrm{e}^{u-Ng_{s}}$, and thus, obtain several plots of the density of
states of the model (\ref{sinh}).

\begin{figure}[htp]
\centering
\includegraphics{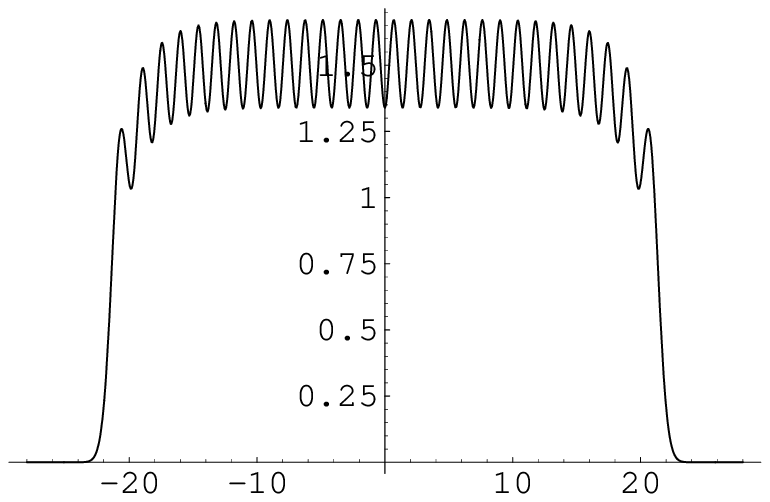}
\caption{$\rho \left( u\right)$ of model (\ref{sinh}), based on (\ref{dens}%
)-(\ref{SW}) with $q=0.5, N=30$}
\end{figure}

Note how the compact support is much bigger than that of a Gaussian
ensemble, for example. Actually, it is of the order $Ng_{s},$ in comparison
with the $\sqrt{N}$ of a Gaussian model. The oscillations that in a Gaussian
model are smoothed out, are here manifest, as can be seen in Fig. 3 for example. 
The number of oscillations coincides with $N$. Crystalline behavior is known to be 
linked with the presence of quantum group symmetries \cite{Tierz2}.

\begin{figure}[htp]
\centering
\includegraphics{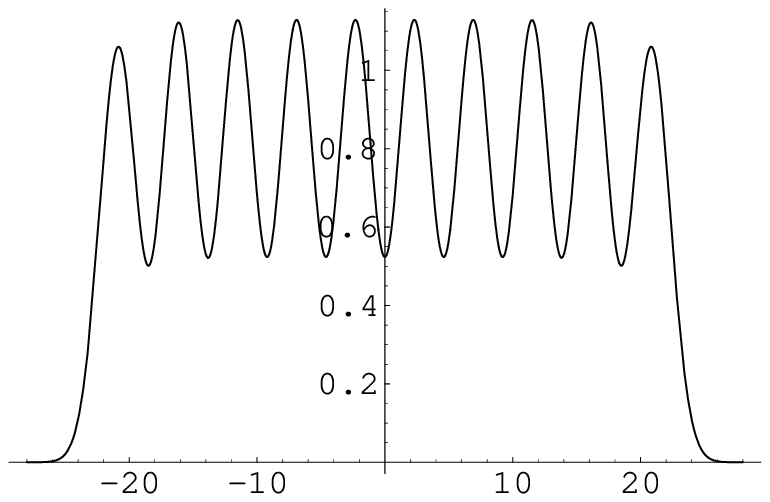}
\caption{$\rho \left( u\right)$ of model (\ref{sinh}), based on (\ref{dens}%
)-(\ref{SW}) with $q=0.3, N=10$}
\end{figure}

With lower values of $q$ the oscillations should be more patent and this is
exactly what happens in the next plot, Figure 4. Note also how the support
grows with $q$. Finally, we go into a higher size where, in contrast to the
Wigner-Dyson case, the amplitude of the oscillatory behavior remains the
same, as can be seen in Figure 5. However, note how keeping the same size
but taking $q$ towards $1$ we are approaching the familiar semi-circle like
behavior. The oscillations considerably decrease in amplitude and so does
the support of the distribution. It is therefore, as shown in Fig. 6, closer to the Wigner-Dyson
case were the oscillations are of smaller amplitude and are essentially
smoothed out when $N$ is very large.

\begin{figure}[htp]
\centering
\includegraphics{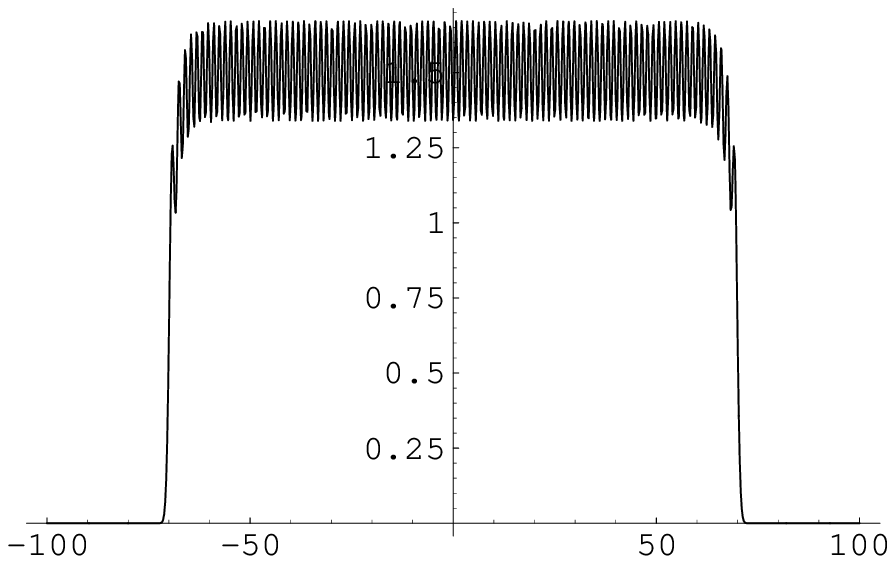}
\caption{$\rho \left( u\right) $ of model (\ref{sinh}), based on (\ref{dens}%
)-(\ref{SW}) with $q=0.5, N=100$}
\end{figure}

\begin{figure}[tph]
\centering
\includegraphics{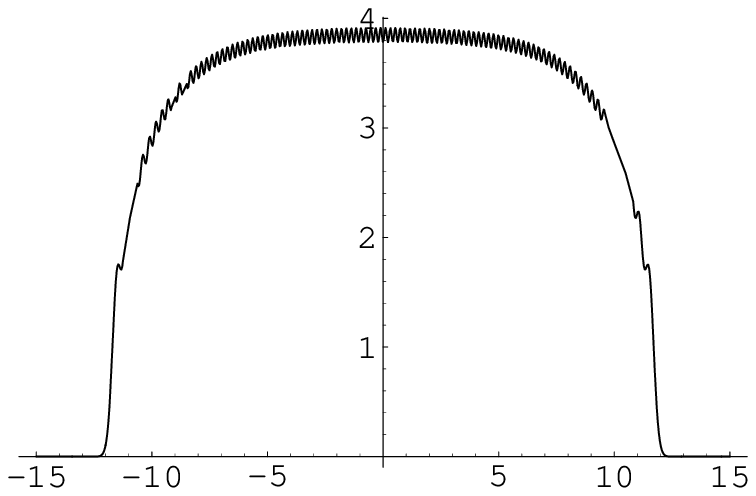}
\caption{$\rho \left( u\right) $ of model (\ref{sinh}), based on (\ref{dens}%
)-(\ref{SW}) with $q=0.9,N=100$}
\end{figure}

\section{Equivalence of effective superpotentials}

Now we would like to further investigate the possible consequences of the
equivalence that we have found. It seems that this equivalence in
Chern-Simons matrix models may be related to recent findings in $\mathcal{N}%
=1$ gauge theory \cite{eq1,eq2,eq3} (see \cite{eqrev} for a review on planar
equivalence of related theories). More precisely, these works show that
different $\mathcal{N}=1$ supersymmetric gauge theories can have an
effective glueball superpotential with exactly the same functional form. In 
\cite{eq2} for example, it is found that the low-energy effective
superpotential of an $\mathcal{N}=1$ $U(N)$ gauge theory with matter in the
adjoint and arbitrary even tree-level superpotential has the same functional
form as the effective superpotential of a $U(N)$ gauge theory with matter in
the fundamental and the same tree-level interactions.

Then, taking into account Dijkgraaf-Vafa \cite{Dijkgraaf:2002fc}, the
effective superpotential of an $\mathcal{N}=1$ supersymetric theory is
computed by a matrix model whose potential is the tree level superpotential
of the gauge theory, then the equivalence of effective superpotentials,
which has been derived employing $\mathcal{N}=1$ techniques -like Konishi
anomaly-, implies the equivalence of the free energies of the corresponding
matrix models.

Note that equivalence of free energies is also one of the mathematical
results obtained in this paper. However, we have discussed Chern-Simons
matrix models. Do these models have any relationship or implication for $%
\mathcal{N}=1$ theories ? Indeed, we have just to recall that at low
energies and in four dimensions, IIA theory compactified on $T^{*}S^{3}$ and
with $N$\ $D6$ branes wrapping $S^{3}$ reduces to $\mathcal{N}=1$ gauge
theory. Therefore, again due to \cite{Dijkgraaf:2002fc}, a Hermitian matrix
model describing CS theory on $S^{3}$ leads to the superpotential of such an 
$\mathcal{N}=1$ theory, which is \cite{Vafa:2000wi}:

\begin{equation}
W=\sum_{n}\left( S+2\pi ni\right) \log \left( S+2\pi ni\right) ^{-N}.
\label{W}
\end{equation}
This role of the Chern-Simons matrix model is already discussed in \cite{Aga}%
. Therefore, this superpotential can be interpreted as the effective
superpotential of an $\mathcal{N}=2$ theory whose tree-level superpotential
is the CS matrix model potential, and we have just seen that there are
infinitely many choices for this potential -of course, once again, all of
them leading to an identical free energy-, so, all of them lead to the same
superpotential, namely to $\left( \ref{W}\right) $.

That is to say, (\ref{W}) as the effective superpotential of an $\mathcal{N}%
=2$ theory whose tree-level superpotential is: 
\begin{equation}
V(M)=k^{2}\log ^{2}M+\log g(M),  \label{V(M)}
\end{equation}
with a $q-$periodic $g(M)$ $=g(qM)$, and $q=\mathrm{e}^{-1/2k^{2}}=\mathrm{e}%
^{-g_{s}}$ as usual$.$ So, regardless the precise form for $g\left( M\right)
,$ we always end up with a superpotential with the same functional form.
Furthermore, recall that the role of (\ref{V(M)}) can also be played by a
discrete potential:

\begin{equation}
V(M)=-\log \left( \sum_{n=-\infty }^{\infty }c^{n}q^{{\frac{n^{2}}{2}}%
+n}\delta (M-cq^{n})\right) ,
\end{equation}
that, dropping constant terms, comes straightforwardly from (\ref
{discreteweight}) and (\ref{disc}).

Finally, just to mention that in \cite{Aganagic:2004js}, a different
Hermitian matrix model for Chern-Simons theory was considered: 
\begin{equation}
V(M)=M^{2}+\frac{1}{2}\sum_{k=1}^{\infty }a_{k}\sum_{s=0}^{2k}\left(
-1\right) ^{s}\binom{2k}{s}\mathrm{TrM}^{s}\mathrm{TrM}^{2k-s}.
\label{hermafroditmonster}
\end{equation}
This cumbersome expression, including double-trace terms, comes from the
transformation: 
\begin{equation}
\prod_{i<j}\left( 2\sinh \frac{u_{i}-u_{j}}{2}\right) ^{2}=\prod_{i<j}\left(
u_{i}-u_{j}\right) ^{2}f\left( u\right) ,  \label{2to1}
\end{equation}
where $f\left( u\right) $ is the logarithm of the second term in (\ref
{hermafroditmonster}) \cite{Aganagic:2004js}. Since this only comes from
manipulations on the repulsion term of the matrix model, it is independent
of the weight function. Thus, employing our result, one can consider the
generalized weight, instead of just the quadratic one and then apply the
method in \cite{Aganagic:2004js}, based on the trick (\ref{2to1}).
Therefore, even if one wants to focus on these type of Hermitian matrix
models, the equivalence result identically holds. That is to say, the
addition of periodic terms, such as the previously discussed (\ref{potential}%
), following the $M^{2}$ term$,$ will not modify the role of (\ref
{hermafroditmonster}), as happens with $g(M)$ in (\ref{V(M)}).

Thus, the arguments and results of this Section are the same regardless the
Hermitian matrix models considered. Another different issue is the
comparison between these Hermitian matrix models, that we shall discuss
elsewhere.

To conclude, note that, while the oscillatory terms may seem at first a bit
awkward, they may be related with the particular form of $\left( \ref{W}%
\right) $. Namely, with the existence of infinitely many domain walls \cite
{Vafa:2000wi}. In addition, these oscillatory terms, as mentioned, seem to
be related to $SL(2,{\mathbb{Z}}),$ which is a crucial ingredient in the
canonical quantization of Chern-Simons theory, and thus in the derivation 
\cite{Marino:2002fk} of the original matrix model.

\section{Conclusions and Outlook}

From \cite{Tierz} and the present work, it can be said that numerous
Chern-Simons quantities can be extracted from a certain \textit{%
coarse-grained} knowledge of either a Gaussian distribution or a log-normal
distribution. This \textit{coarse-grained }knowledge, is nothing else than a
suitable combination of the set of positive integer moments of the
distribution. This particular combination of the integer moments is the one
given by the orthogonal polynomials. But it turns out that the matrix models
have such a fluctuating weight function that this coarse-grained information
-the moments- no longer fixes the weight function uniquely, in sharp
contrast with polynomial potentials, for example. Therefore, one has the
freedom of modifying the weight function, while leaving intact certain
amount of information. This reduced information is the only one required by
the matrix model, and consequently by Chern-Simons theory. We have shown the
rather large extent to which different superpotentials can be equivalent.
Mathematically, both the discrete case and the oscillatory -especially the
ones with very low differentiability- seem interesting on their own. Note
that the most extreme cases are matrix models with fractal potentials that,
nevertheless, give rise to physically meaningful objects like Chern-Simons
partition functions.

Regarding possible avenues for further research, discrete matrix models have
been poorly studied in gauge theory, at least in comparison with continuous
models. The technology available in these type of models is already quite
formidable and, in addition, many remarkable combinatorial properties of
these models are known \cite{Johansson}, so their study may prove useful in
relationship with concepts relevant to topological string theory, like
crystal melting. In addition, as discussed above, our methods provide a
direct understanding of why at the level of the matrix models Chern-Simons
theory and 2d $q$YM are equivalent. It also gives a concrete computational
scheme where one can straightforwardly extend this equivalence between
continuous and discrete models to other cases, namely, to more general
manifolds like Seifert homology spheres, and generic torus link invariants.
It would be interesting to see whether the discrete model that one gets is
still reproduced by a BF-type of theory, and on which manifold. Hopefully,
the equivalence between the continuous and the discrete matrix models can be
applied to compute large $N$ limits of 2dYM and topological strings on more
general manifolds as well. Needless to say, our results apply also to
orthogonal and symplectic groups, so one could consider Calabi-Yau
orientifolds.

Interestingly enough, it seems that this non-uniqueness also leads to the
equivalence of effective superpotentials in $\mathcal{N}=1,$ through a
rather different route from \cite{eq1,eq2,eq3}. A further understanding of
this seems desirable. Another very much related question would be to
understand if the precise way in which this non-uniqueness occurs in the
matrix models is related to the specific form of the superpotential $\left( 
\ref{W}\right) $, found in \cite{Vafa:2000wi}.

\bigskip

\textbf{Acknowledgments}

\medskip

We are grateful to Riccardo Argurio, Edouard Br\'{e}zin, Alexander Gorsky,
Michel Lapidus, Juan Luis Ma\~{n}es, Marcos Mari\~{n}o, Ingo Runkel, Stefan
Theisen and Jean-Bernard Zuber for discussions and correspondence.

\newpage \appendix

\section{Notation and Chern-Simons quantities}

We give here some information about the Chern-Simons quantities that appear
in the text, mainly in $\left( \ref{20}\right) $. For more information, see 
\cite{Marino:2002fk} and references therein. To understand the origin of
other quantities in $\left( \ref{20}\right) $, one has to take into account
the constructions of Seifert homology spheres from surgery. Seifert homology
spheres can be constructed by performing surgery on a link $\mathcal{L}$ in $%
\mathbf{S}^{3}$ with $n+1$ components, consisting on $n$ parallel and
unlinked unknots together with a single unknot whose linking number with
each of the other $n$ unknots is one. The surgery data are $p_{j}/q_{j}$ for
the unlinked unknots, $j=1,\cdots ,n$, and 0 on the final component. $p_{j}$
is coprime to $q_{j}$ for all $j=1,\cdots ,n$, and the $p_{j}$'s are
pairwise coprime. After doing surgery, one obtains the Seifert space $M=X({%
\frac{p_{1}}{q_{1}}},\cdots ,{\frac{p_{n}}{q_{n}}})$. This is rational
homology sphere whose first homology group $H_{1}(M,\mathbf{Z})$ has order $%
|H|$, where 
\begin{equation}
H=P\sum_{j=1}^{n}{\frac{q_{j}}{p_{j}}},\,\,\,\,\,\mathrm{and}%
\,\,\,\,P=\prod_{j=1}^{n}p_{j}.  \label{orderh}
\end{equation}
Another topological invariant that will enter the computation is the
signature of $\mathcal{L}$, which turns out to be: 
\begin{equation}
\sigma (\mathcal{L})=\sum_{i=1}^{n}\mathrm{sign}\biggl( {\frac{q_{i}}{p_{i}}}%
\biggr) -\mathrm{sign}\biggl( {\frac{H}{P}}\biggr).  \label{signa}
\end{equation}
For $n=1,2$, Seifert homology spheres reduce to lens spaces, and one has
that $L(p,q)=X(q/p)$. For $n=3$, we obtain the Brieskorn homology spheres $%
\Sigma (p_{1},p_{2},p_{3})$ (in this case the manifold is independent of $%
q_{1},q_{2},q_{3}$). In particular, $\Sigma (2,3,5)$ is the Poincar\'{e}
homology sphere. Finally, the Seifert manifold $X({\frac{2}{-1}},{\frac{m}{%
(m+1)/2}},{\frac{t-m}{1}})$, with $m$ odd, can be obtained by integer
surgery on a $(2,m)$ torus knot with framing $t$. Note that in $\left( \ref
{20}\right) $ the weight and root lattices of $G$ are denoted by $\Lambda _{%
\mathrm{w}}$ and $\Lambda _{\mathrm{r}}$, respectively.

Finally, there is a phase factor in $\left( \ref{20}\right) $ that comes
from the framing correction, that guarantees that the resulting invariant is
in the canonical framing for the three-manifold $M$. Its explicit expression
is: 
\begin{equation}
\phi =3\mathrm{sign}\left( \frac{H}{P}\right) +\sum_{i=1}^{n}12s\left(
q_{i},p_{i}\right) -\frac{q_{i}}{p_{i}},
\end{equation}
where $\sigma (\mathcal{L})$ is again the signature of the linking matrix of 
$\mathcal{L}$ and $s(p,q)$ is the Dedekind sum 
\begin{equation}
s(p,q)={\frac{1}{4q}}\sum_{n=1}^{q-1}\cot \Bigl({\frac{\pi n}{q}}\Bigr)\cot %
\Bigl({\frac{\pi np}{q}}\Bigr).
\end{equation}

\end{document}